\documentclass{ws-mpla}
\begin{document}

\markboth{Nijo Varghese and V. C. Kuriakose} {Late-time evolution of Dirac field around Schwarzschild-quintessence black hole}
%%%%%%%%%%%%%%%%%%%%% Publisher's Area please ignore %%%%%%%%%%%%%%
\catchline{}{}{}{}{}
%%%%%%%%%%%%%%%%%%%%%%%%%%%%%%%%%%%%%%%%%%%%%%%%%%%%%%%%%%%%%%%%%%%
\title{LATE-TIME EVOLUTION OF DIRAC FIELD AROUND SCHWARZSCHILD-QUINTESSENCE BLACK HOLE}

\author{\footnotesize NIJO VARGHESE\footnote{Presently at Sacred Heart College, Kochi, 682013, India}}

\address{Department of Physics, Cochin University of Science and Technology\\ Kochi, 682022, 
India\\
nijovarghese@cusat.ac.in}

\author{V C KURIAKOSE}
\address{Department of Physics, Cochin University of Science and Technology\\ Kochi, 682022, India
\\ vck@cusat.ac.in}

\maketitle
% \pub{Received (Day Month Year)}{Revised (Day Month Year)}

\begin{abstract}
The late-time evolution of Dirac field around spherically symmetric black hole surrounded by quintessece is studied numerically. 
Our results show, for lower values of the quintessence state parameter $\epsilon$, Dirac field decays as power-law tail 
but with a slower decay rate than the corresponding Schwarzschild case. 
But for $\epsilon<-1/3$, all the $\ell$-poles of the Dirac field give up the power-law decay form and relax to a constant residual 
field at asymptotically late times. The value of this residual field for which the field settles down varies on different surfaces. 
It has the lowest value on the black hole event horizon, increases as the radial distance increases and maximizes on the cosmological horizon.

\ccode{PACS Nos.: 95.36.+x, 04.70.Bw, 03.65.Pm} 
\end{abstract}

\section{Introduction}
\label{intro}
According to the black hole \textquotedblleft {\it no-hair theorem}\textquotedblright\, a black hole formed by the gravitational collapse of 
a charged rotating star, will rapidly relaxes to the stationary state, characterized by three quantities, its mass, charge and angular 
momentum\cite{NHT}. Any other \textit{hair} will disappear after the collapsing body settles down to 
its stationary state. Configurations violating the generalized no-hair conjecture were also presented including black holes dressed 
with Yang-Mills\cite{hairYangMills}, Skyrme\cite{hairSkyrme} and dilaton\cite{hairDilaton} fields, 
in various combinations with Higgs fields\cite{YangMillsHiggs}. 
Even though most of these black holes are found to be unstable, there are few stable solutions also. 
By analyzing the stability of a black hole solution of the Einstein-Yang-Mills equations in the framework of small
time-dependent perturbations it was shown that there is at least one exponentially growing radial mode with the correct boundary
conditions at the horizon and at infinity\cite{hairinstable1}. 
It was proven that there are unstable modes of the Bartnik-McKinnon soliton and
the non-abelian black hole solution of the Einstein-Yang-Mills theory for the gauge group SU(2)\cite{hairinstable2,hairinstable3}. 
The manner and rate with which the hairs of the black hole decay is thus an important question. 

It was Richard Price\cite{price1972} who, making a perturbative analysis of the 
collapse of a nearly spherical star, showed that for a field with spin, s, any radiative multipole 
($\ell\geq s$) gets radiated away completely, in the late stage of collapse. Further, he showed that at late 
times the field dies out with a power-law tail $t^{-(2\ell+p+1)}$, where $p=1$, if the multipoles were initially 
static and $p=2$ otherwise. Price's theorem was verified for various field perturbations around different black hole 
spacetimes in asymptotically flat spacetimes\cite{gundlach1994,hod1998a,hod1998b,NV2011,koyama,konoplya2007,cordoso2003,abdalla2005,schen2008}.

%========================================

In the past two decades there have been growing observational evidences\cite{perlmutter1999,riess1998} which established clearly that 
our universe is expanding in an accelerated pace, indicating the presence of some mysterious form of repulsive energy called dark energy. 
In order to explain the nature of dark energy, several models have been proposed. The simplest option being Einstein's cosmological
constant($\Lambda$), which has a constant equation of state with state parameter, $\epsilon=-1$, but it needs extreme fine tuning to account
for the observations\cite{weinberg}. Models were proposed, replacing $\Lambda$ with a dynamical,
time-dependent and spatially inhomogeneous component now called quintessence, which can have an equation of state, 
$-1\leq\epsilon\leq-1/3$\cite{peebles,ratra,caldwell}. Quintessence hypothesis is found to fit current observations and more 
precise measurements may separate the two models in future. So it is interesting to check Price's theorem for a black hole with nonflat asymptotes.

For black holes in asymptotically de Sitter spacetimes, a generalization of Price's theorem was presented in\cite{chambers}, for massless 
small fluctuations. Later the classical black hole no-hair theorems were extended to spacetimes endowed with a positive $\Lambda$ 
for different fields\cite{bhattacharya2007}. In the perturbative study of Price's theorem, the late-time decay is determined by the 
asymptotic curvature of the spactime\cite{ECSching}. The studies on the time evolution of scalar, electromagnetic and gravitational 
perturbations, propagating on black hole with asymptotically de Sitter like spacetime revealed the existence of an exponentially 
decaying tails at late times contrasting the power-law tails in asymptotically flat situation\cite{brady1997,brady1999,molina,NV2013}. 

%%%------------------------------------Dirac--------------------------------

The aim of this paper is to study the late-time evolution of Dirac field around a black hole whose asymptotes are determined by quintessence.
For black hole with asymptotically flat spacetimes, the late-time behavior of Dirac field is well understood
\cite{jjing2004,jjing2005,jjing2006}. The intermediate and late-time behavior of massive Dirac field, in the static spherically symmetric general black
hole spacetime, is studied in \cite{moderski2008}. The late-time behavior of a massive Dirac field in the background of 
dilaton and brane-world black hole solutions is investigated in \cite{gibbson2008a,gibbson2008b}. It is revealed that for black hole 
in flat spacetimes, the long-lived oscillatory tail of massive Dirac field, decays as $t^{-5/6}$.

For Schwarzschild-de Sitter(SdS) black holes, the quasinormal modes(QNMs) of decay had been calculated for fields of different spin, 
including Dirac field \cite{zhidenko2004}. The QNMs of Dirac field around black holes surrounded by quintessence were calculated 
in\cite{zhang2009,wang2010}. But the late-time behavior of Dirac fields in these spacetimes is not clear. 
Recently, a proof of the no-hair theorem corresponding to perturbative massive spin-1/2 fields for stationary axisymmetric 
de Sitter black hole is presented\cite{bhattacharya2012}. So it is interesting to see how does Dirac field evolve in a spacetime 
in which the asymptotic structure is altered by the quintessence field.

The rest of the paper is organized as follows. In Sect.\ref{sec2} we introduce the master wave equation for Dirac field 
perturbations around black hole surrounded by quintessence. The numerical method used to study the time evolution is 
explained in Sect.\ref{sec3} and the results are presented. The conclusion and discussions are given in Sect.\ref{sec4}.

\section{Dirac field around black hole surrounded by quintessence}
\label{sec2}
The exact solution of Einstein's equation for a static spherically symmetric black hole surrounded by the quintessential matter,
under the condition of additivity and linearity in energy momentum tensor, was found in\cite{kiselev},

\begin{equation}
\ ds^{2}=-f(r)dt^{2}+f(r)^{-1}dr^{2}+r^{2}(d\theta ^{2}+\sin^{2}\theta d\phi ^{2}), \label{metric}
\end{equation}

where $f(r)=\left(1-\frac{2M}{r}-\frac{c}{r^{3\epsilon+1}}\right)$, $M$ is the mass of the black hole, $\epsilon$ is the 
quintessential state parameter and $c$ is the normalization factor which depends on the density
of quintessence as, $\rho_{q}=\frac{-c}{2}\frac{3\epsilon }{r^{3(1-\epsilon )}}$.
Various properties of this black hole solution were studied in the past(See Ref. \refcite{tharanath2013,tharanath2014} for instance).
%%%%%----------------------------------------------------
It is difficult to analyze the perturbation in the above metric for arbitrary values of the parameter $\epsilon$.
For our study we take three special cases of the quintessence parameter $\epsilon=-1/3,-2/3$ and -1, so that one can get 
simple expression for the radial tortoise coordinate in terms of the horizons and surface gravity associated with the horizons\cite{NV2013} 
and the calculations become viable.

For $\epsilon=-1/3$, the black hole event horizon is located at $r_{e1}=2M/(1-c)$. When $\epsilon=-2/3$, in addition to the black hole 
event horizon at $r=r_{e2}$ the spacetime possesses a cosmological horizon at $r=r_{c2}$, with $r_{e2}<r_{c2}$. 
The surface gravity associated with the horizons at $r=r_{i}$, is defined by $\kappa_{i}=\frac{1}{2}|df/dr|_{r=r_{i}}$ and we get,
\begin{equation}
\kappa_{e2}=\frac{c(r_{c2}-r_{e2})}{2r_{e2}}, \quad\quad  \kappa_{c2}=\frac{c(r_{e2}-r_{c2})}{2r_{c2}}. 
\end{equation}

The extreme case of quintessence, $\epsilon=-1$, corresponds to the SdS spacetime. 
The surface gravity at the event horizon, $r=r_{e3}$ and the cosmological horizon $r=r_{c3}$, are given by, 

\begin{equation}
\kappa_{e3} = \frac{c(r_{c3}-r_{e3})(r_{e3}-r_{0})}{2r_{e3}}, \quad\quad \kappa_{c3} = \frac{c(r_{c3}-r_{e3})(r_{c3}-r_{0})}{2r_{c3}},
\end{equation}

where the third root of the polynomial equation $f(r)=0$, $r_{0}=-(r_{e3}+r_{c3})$, with $r_{0}<r_{e3}<r_{c3}$.

%====================================================================================================================

The Dirac equation for a massless field in spacetime $g_{\mu\nu}$, specified by Eq.(\ref{metric}) has the form\cite{birrell},

\begin{equation}
i\gamma ^{a}e_{a}^{\mu }(\partial _{\mu }+\Gamma _{\mu })\Psi =0, \label{diraceqn}
\end{equation}

where $\gamma ^{a}$ are the Dirac matrices, $\Gamma _{\mu }$ are the spin connections and $e_{a}^{\mu }$ are the tetrad.

The radial part of the above perturbation equations can be decoupled from their angular parts and reduced to the form\cite{HTcho2003,zhang2009},

\begin{equation}
\left(-\frac{\partial^{2}}{\partial t^{2}}+\frac{\partial^{2}}{\partial r^{2}_{*}}\right)\Psi_{\ell}(t,r)=-V_{\pm}(r)\Psi_{\ell}(t,r),
\label{waveqn}
\end{equation}

where the tortoise coordinate is defined by, $dr_{*}=(1/f)dr$ and the effective potentials are given by,

\begin{equation}
V_{\pm} = \frac{|k|\sqrt{f}}{r^{2}}\left[ |k|\sqrt{f}\pm\frac{r}{2}\frac{\partial f}{\partial r}\mp f\right], \label{potentials}
\end{equation}
were $k$ is a positive or a negative nonzero integer related to the total orbital angular momentum by $k=\ell +1$ for $(+)$ sign 
and $k=\ell$ for $(-)$ sign. The potentials, $V_{+}$ and $V_{-}$ are the super symmetric partners and
give same spectra\cite{anderson1991}, so we choose $V_{+}$ by omitting the subscripts. 

\begin{figure}[h]
\centering
\includegraphics[width=0.7\columnwidth]{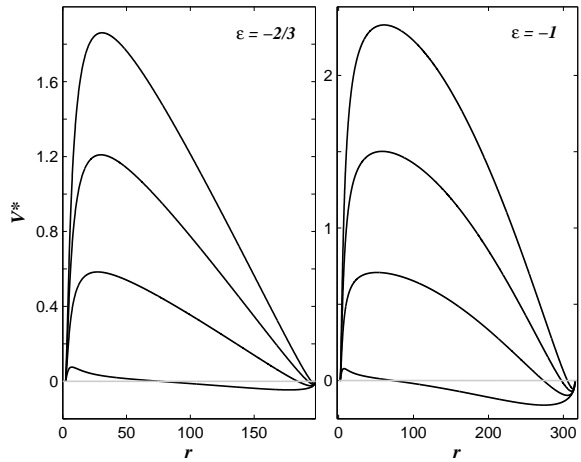}%
\caption{Plots of effective potentials experienced by the Dirac field for quintessence parameters $\epsilon = -2/3$ with $c=10^{-2}/2$ and 
$\epsilon = -1$ with $c=10^{-5}$. Curves from bottom to top are for $\ell=0,1,2$ and 3 modes. The potential is scaled as 
$V^{*}=V(r_{e}-r)^{2}(\ell+1/2)$, to enhance the features at large r. Potentials have a negative dip near cosmological horizon.} 
\label{potplot}
\end{figure}

For $\epsilon=-1/3$, the effective potential is positive definite for $r_{*}\in[-\infty,+\infty]$ and has a potential 
barrier near the event horizon but vanishes asymptotically as  $r_{*}\rightarrow\pm\infty$. 
As the parameter $\epsilon$ decreases below -1/3, a cosmological horizon is created by the quintessence. Figure \ref{potplot} shows effective 
potentials for $\epsilon=-1$ and -2/3. In these cases, after a barrier nature near the event horizon, the effective potentials 
for all modes, vanish at some $r_{*}=r_{*}^{0}$ and there after form a negative well in the range $r_{*}^{0}<r_{*}<+\infty$. 
This behavior of Dirac field is in contrast with other fields. For scalar field, even if the $\ell=0$ mode shows the negative dip 
in the potential, all other higher modes have a positive value for the potential between the horizons\cite{NV2013,brady1997}.

\section{Numerical integration and results}
\label{sec3}
The complex nature of the potentials makes it difficult to obtain the exact solutions of Eq.(\ref{waveqn}) and we have to tackle the
problem by numerical methods. A simple and efficient method to study the evolution of field were developed in\cite{gundlach1994},
 after recasting the wave equation, Eq.(\ref{waveqn}), in the null coordinates, $u=t-r_{*}$ and $v=t+r_{*}$ as,
 
\begin{equation}
-4\frac{\partial^{2}}{\partial u \partial v}\Psi(u,v)=V(u,v)\Psi(u,v) \label{waveqn2}
\end{equation}

and using the following discretization,

\begin{equation}
\Psi_{N}=\Psi_{W}+\Psi_{E}-\Psi_{S}-\frac{h^{2}}{8}V(S)(\Psi_{W}+\Psi_{E})+O(h^{4}).
\end{equation}

The numerical integration is performed on an uniformly spaced grid with points, $N(u+h,v+h)$, $W(u+h,v)$, $E(u,v+h)$ and $S(u,v)$ 
forming a null rectangle with an overall grid scale factor of $h$. The tortoise coordinates are inverted using Newton-Raphson 
method\cite{NV2013}. The field is scaled as, $\phi=\psi/r$ and the evolution is monitored on different surfaces viz., 
\begin{enumerate}
 \item the cosmological horizon (approximated by the null surface, $v=v_{max}$),
 \item the black hole event horizon(approximated by the null surface, $u=u_{max}$) and
 \item different null surfaces of fixed radius, $r_{*}=K$.
\end{enumerate}

%=============================================================================RESULTS================================================

\begin{figure}[h]
\centering
\includegraphics[width=0.47\columnwidth]{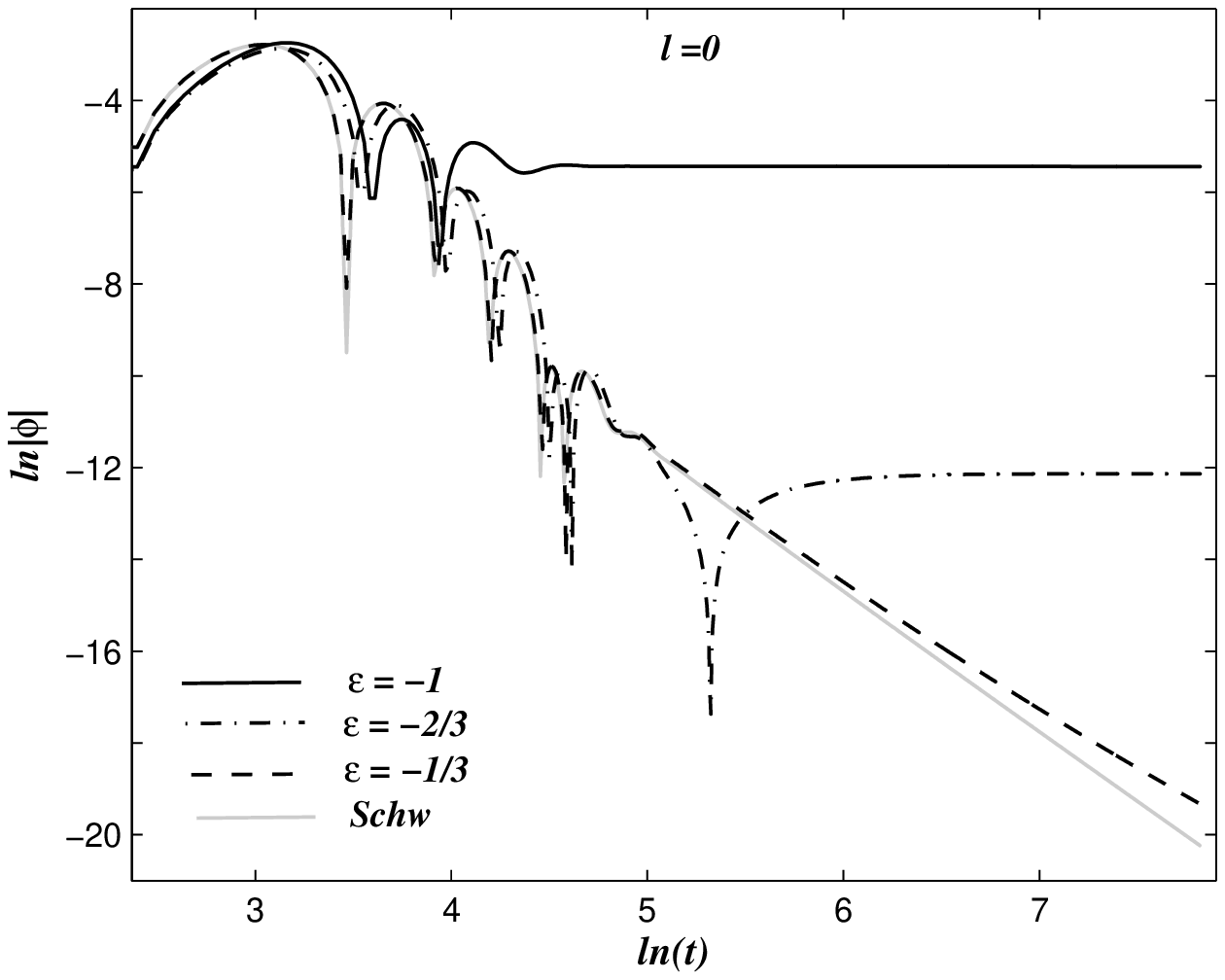}
\includegraphics[width=0.47\columnwidth]{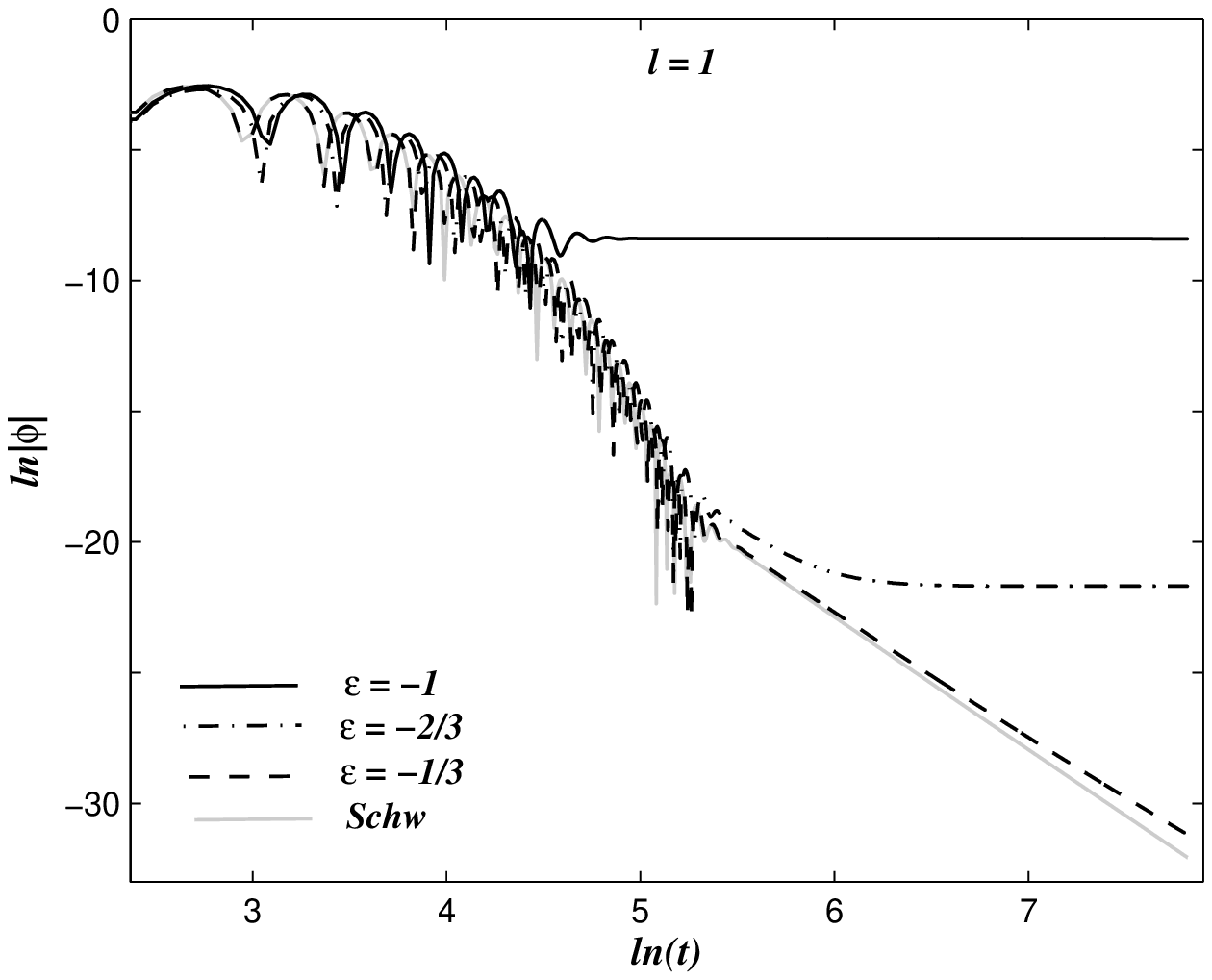}
\caption{Log-log graph of the evolution of Dirac field in a quintessence filled black hole spacetime with $c=10^{-2}/2$, in comparison 
with that in the pure Schwarzschild spacetime, evaluated at $r*=10$. $\ell=0$ and $\ell=1$ modes for $\epsilon=-1/3,-2/3$ and -1.} 
\label{combined}
\end{figure}

Figure \ref{combined} shows the evolution profile of the Dirac field around the black hole in a quintessence filled universe 
along with that in the pure Schwarzschild spacetime. Monopole and dipole fields are shown. We observe that the evolution shows 
deviations from the Schwarzschild case, after initial transient phase. The damped oscillation(QNM) phase and the 
late-time tail of decay in the final phase(these two phases depend only on the characteristics of the background black hole 
spacetime\cite{nollert1999}) show the signature of the quintessence. We observe that the field decays slowly in 
the QNM phase, if quintessence is present as it was shown in\cite{zhidenko2004,zhang2009}, using the WKB method. The QNM phase is 
followed by the regime of late-time tails of field decay. 

It is well known that, the field has the power-law decay in the pure Schwarzschild spacetime, represented by the straight lines in a 
log-log plot. Power-law decay is observed for the $\epsilon=-1/3$ case of quintessence, but with slightly slower 
decay rate than the corresponding Schwarzschild tail. For $c=10^{-2}/2$, we get $\phi\sim t^{-(2\ell+2.7)}$, a slower decay than 
the $\phi\sim t^{-(2\ell+3)}$ of the pure Scharzschild case.

\begin{figure}[h]
\centering
\includegraphics[width=0.6\columnwidth]{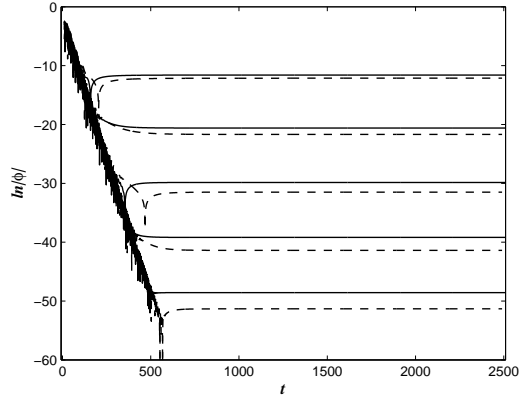}
\caption{Evolution of Dirac field extracted at $r*=10$ for the $\epsilon=-2/3$ case with $c=10^{-2}/2$(solid curves) and $\epsilon=-1$ case with 
$c=10^{-5}$(dashed curves). In each case, curves from top to bottom are for $\ell=0,1,2,3$ and 4.} %
\label{DE123L}
\end{figure}

Figure \ref{DE123L} shows the evolution profile for the $\epsilon=-2/3$ and $-1/3$ cases. At late times, the field does not decay for 
these cases. All the $\ell$-poles of field relaxes to a constant residual field, $\phi_{0}$ at asymptotic late times. This behavior of 
Dirac field is in contrast with other spin fields, for which an exponential decay was observed\cite{brady1997,molina,NV2013}.
Similar behavior was observed for the monopole of the scalar field but all the $\ell>0$ modes were found to be exponentially decaying.
The nature of Dirac field is little surprising and it strengthens the dependence of the unusual negative dip in the potential near cosmological 
horizon and the relaxation of the field to a constant value. In order to verify that the residual static field is not a numerical 
error  we checked the convergence of the code by decreasing the grid space and can find a good convergence as $h\rightarrow 0$.

\begin{figure}[h]
\centering
\includegraphics[width=0.47\columnwidth]{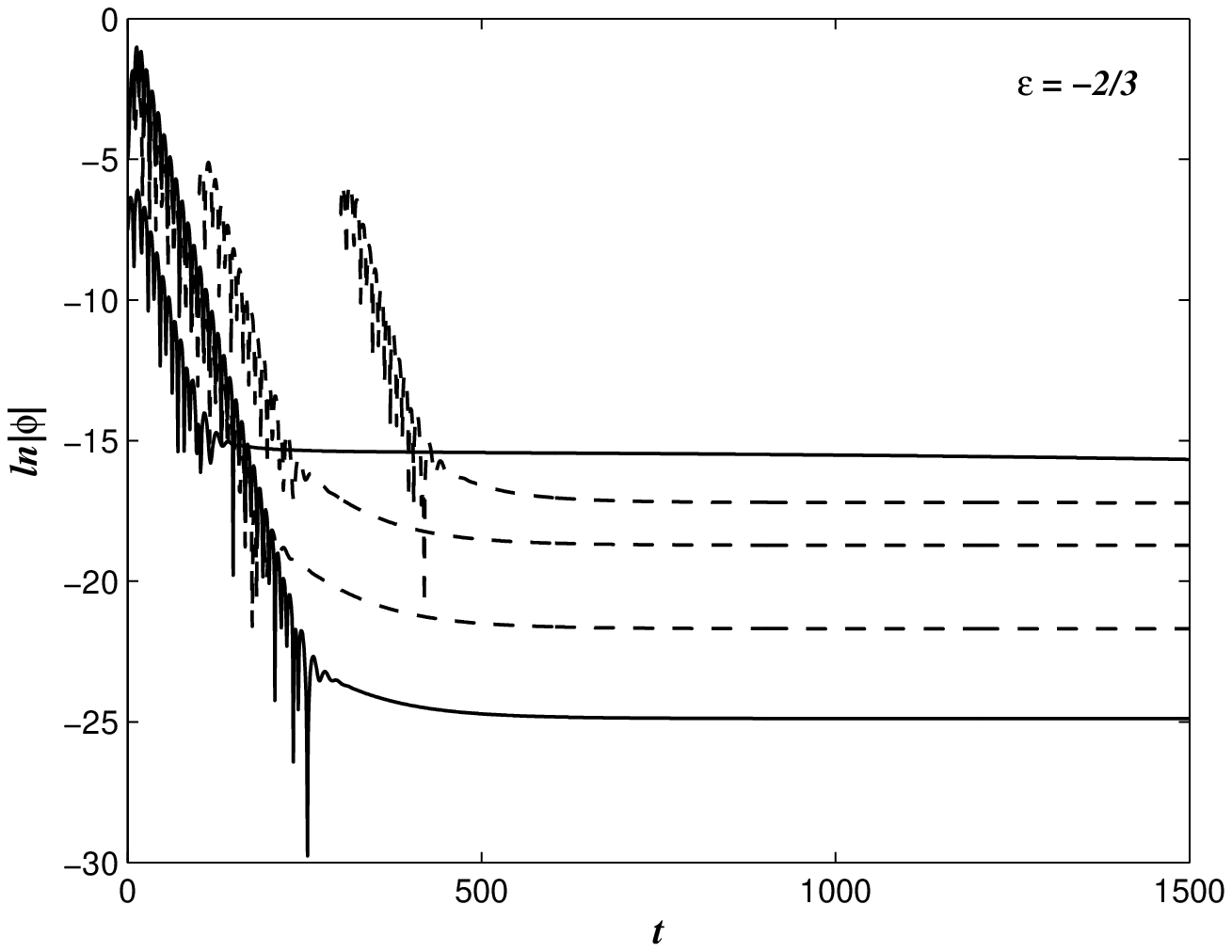}%
\includegraphics[width=0.47\columnwidth]{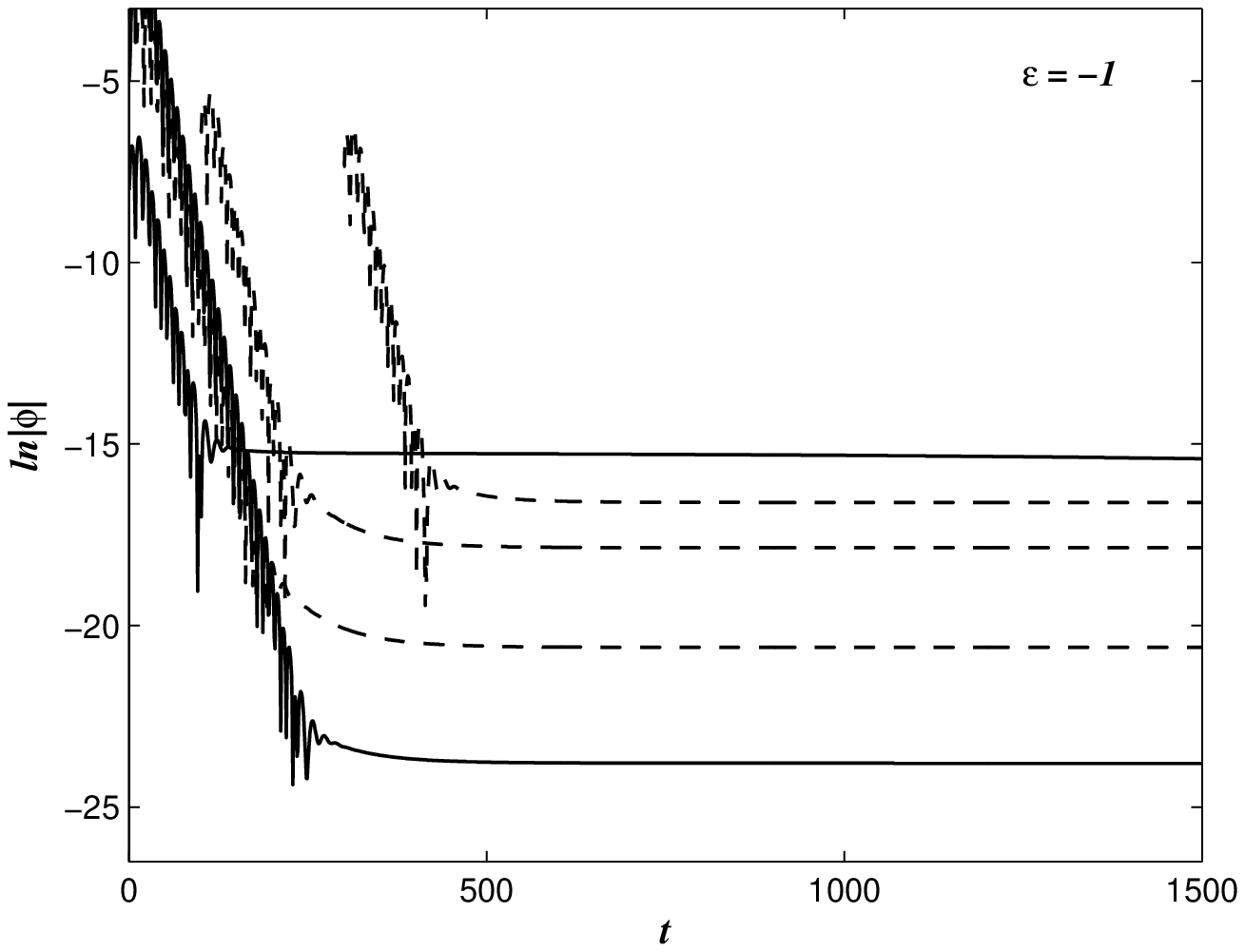}
\caption{The decay of $\ell=2$ mode of Dirac field on different surfaces. Solid curves represent the field on event horizon(bottom) 
and cosmological horizon(top). Dotted curves from bottom to top correspond to the field on surfaces at $r*=10, 100$ and 300.}
\label{E23Surf4}
\end{figure}

To confirm that the behavior of Dirac field is not an artifact of the particular location, we monitor the evolution of the field on 
different null surfaces of constant radius and on the event and cosmological horizons. Figure \ref{E23Surf4} shows the decay of 
$\ell=2$ mode of Dirac field on the black hole event horizon, cosmological horizon and three surfaces of fixed radius, $r*=10, 100$ and 300.
The constant asymptotic value of the Dirac field, $\phi_{0}$, varies from the black hole event horizon to the cosmological horizon. 
The $\phi_{0}$ has a lowest value on the event horizon, increases as radial position goes farther, and has the highest value on the 
cosmological horizon.

\begin{figure}[h]
\centering
\includegraphics[width=0.7\columnwidth]{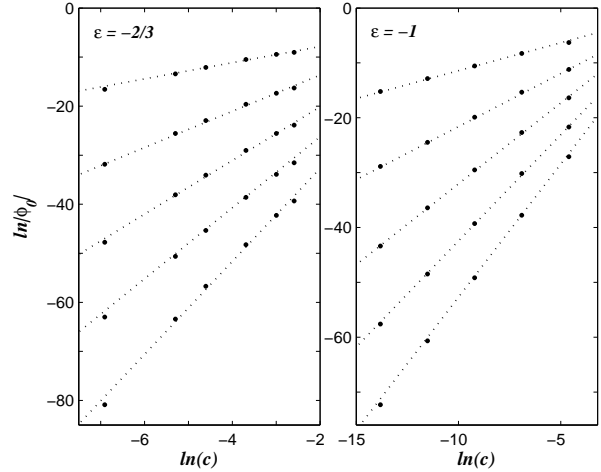}
\caption{The asymptotic value of the Dirac field along event horizon, $\phi_{0}$ plotted against $c$, in logarithmic scale.
Dotted lines represent a linear fit. Curves from top to bottom are for $\ell=0,1,2,3$ and 4.} %
\label{DE23DL}
\end{figure}

The dependence of the asymptotic residual field, $\phi_{0}$, on the parameter c, is shown in Fig.\ref{DE23DL}, in logarithmic scale. 
For $\epsilon = -1$, a least square fit for $ln|\phi_{0}| = m\ ln(c) + c1$, gives the slopes 0.973, 1.933, 2.942, 3.915 and 4.921 
for $\ell = 0,1,2,3$ and 4, respectively. The y intercepts are 1.921, 4.288, 7.470, 8.646 and 10.716. For $\epsilon = -2/3$, we get 
the slops 1.768, 3.630, 5.559, 7.330 and 9.625 and y intercepts -0.445, -0.423, -0.519, -0.897 and -0.744, for $\ell = 0,1,2,3$ and 4, 
respectively. These results suggest that,  

\begin{eqnarray}
\phi_{0} & \sim & c^{(\ell+1)}, \qquad \ \ \ \ \ \ for \quad \epsilon = -1, \nonumber \\
\phi_{0} & \sim & c^{1.782(\ell+1)}, \qquad for \quad \epsilon = -2/3.
\label{dir}
\end{eqnarray}

\section{Conclusions and discussions}
\label{sec4}

The paper studies the evolution of Dirac field perturbation, particularly the late-time behavior, around a black hole spacetime surrounded 
by quintessence. The quintessence equation of state, $\epsilon$, plays a dramatic role in the late-time decay of the Dirac field. 
For $\epsilon=-1/3$ the late-time decay follows a power-law form, but with a lower decay rate than the corresponding Schwarzschild case.
As the value of the quintessential parameter $\epsilon$, decreases the cosmological horizon forms and a negative dip appears in the 
effective potential near the cosmological horizon. As it seems to be the consequence of the peculiar behavior of potential, for 
$\epsilon=-2/3$ and -1, the Dirac field does not decay to zero, but relaxes to a constant residual field, at late times. The values of 
the residual field is determined by the values of the parameter $\epsilon$ and c.

The asymptotic value of the Dirac field varies on different surfaces of constant radius. It has the lowest value on the black hole 
event horizon, increases as the radial distance increases and maximizes on the cosmological horizon. This behavior of Dirac field seems 
to be odd comparing with other spin field perturbations, where all the $\ell>0$ modes of the field decay exponentially. 
Price's original work demonstrates that there can be no static solution to the scalar wave equation that are well behaved at infinity 
and at black hole event horizon. Even though the $\ell=0$ mode of the scalar field in the SdS spacetime is observed to settle down to a constant 
asymptotic value, it relaxes to the \textit{same} constant value on all the surfaces. It can be argued that the constant field 
does not carry any stress energy tensor and it is equivalent to vanishing of the hair. But the behavior of Dirac field is rather intriguing 
since all the $\ell$ modes of the field have non zero value at late times and it varies on different surfaces. Our study indicates that 
there may be static solutions for Dirac field for all $\ell$ for black holes with de-Sitter like asymptotes. 
The presence non decaying wave tails at late times may lead to the instability of Cauchy horizons inside charged and rotating black holes
and the strong cosmic censorship has to be revisited\cite{brady4} for these spacetimes. 
Further detailed numerical and analytical studies of the spin-1/2 fields around black holes in an expanding universe are, hence, call forth.

\section*{Acknowledgments}
The authors thank Alexander Zhidenko for useful discussions. 
NV wishes to thank the University Grants Commission(UGC), New Delhi, for financial assistance under
DSKPDF scheme. VCK is thankful to UGC, New Delhi for financial support through a Major Research Project 
and wishes to acknowledge Associateship of IUCAA, Pune, India.


\begin{thebibliography}{00}

\bibitem{NHT}Israel W 1967 \textit{Phys. Rev.} \textbf{164} 1776; 
             Carter B (1971) \textit{Phys. Rev. Lett.} \textbf{26} 331; 
             Wald R M 1971 \textit{Phys. Rev. Lett.} \textbf{26} 1653; 
             Hawking S W 1991 \textit{Commun. Math. Phys.} \textbf{25} 152
             \bibitem{hairYangMills} Bizon P, 1990 \textit{Phys. Rev. Lett.} \textbf{64} 2844;
                        Kuenzle H P and Masood A K M 1990 \textit{J. Math. Phys.} \textbf{31} 928;
                        Volkov M S and Galtsov D V \textit{JETP Lett.} \textbf{50} 346.
\bibitem{hairSkyrme}Droz S, Heusler M and Straumann N 1991 \textit{Phys. Lett. B} \textbf{268} 371; 
                    Heusler M, Straumann N and Zhou Z-H 1993 \textit{Helv. Phys. Acta} \textbf{66}  614.
                    % Heusler M., Droz S. and Straumann N., Phys. Lett. B 271: 61  (1991); Phys. Lett. B 285 : 21 (1992); % Luckock H. and Moss I. (1986), Phys. Lett. B 176: 34
\bibitem{hairDilaton} Lavrelashvili G and Maison D 1993 \textit{Nucl. Phys. B} \textbf{410} 407
\bibitem{YangMillsHiggs} Breitenlohner P, Forgacs P and Maison D 1992 \textit{Nucl. Phys. B} \textbf{383} 357; 
                        Greene B R, Mathur S D and O'Neill C M 1993 \textit{Phys. Rev. D} \textbf{47} 2242.
\bibitem{hairinstable1} Straumann N and Z H Zhou 1990 \textit{Phys. Lett. B} \textbf{243} 33.
\bibitem{hairinstable2} Lavrelashvili G and Maison D 1995 \textit{Phys. Lett. B} \textbf{343} 214.
\bibitem{hairinstable3} Volkov M S, Brodbeck O, Lavrelashvili G and Straumann N 1995 \textit{Phys. Lett. B} \textbf{349} 438.

%%-------------------------------Price's theorem--------------------------
\bibitem{price1972} Price R H 1972 \textit{Phys. Rev. D} \textbf{5}; 2419, \textbf{5}, 2439
\bibitem{gundlach1994} Gundlach C, Price R H and Pullin J 1994 \textit{Phys. Rev. D} \textbf{49}, 88
\bibitem{hod1998a} Hod S and Piran T 1998 \textit{Phys. Rev. D} \textbf{58}, 024017; \textbf{58}, 024018; \textbf{58} 024019
\bibitem{hod1998b} Hod S and Piran T 1998 \textit{Phys. Rev. D} \textbf{58} 044018
\bibitem{NV2011} Varghese N and Kuriakose V C 2011 \textit{Gen. Relativ. Gravit.} \textbf{43} 2757
\bibitem{koyama} Koyama H and Tomimatsu A 2001 \textit{Phys. Rev. D} \textbf{63} 064032
\bibitem{konoplya2007} Konoplya R A, Zhidenko A and Molina C 2007, \textit{Phys. Rev. D} \textbf{75} 084004 
\bibitem{cordoso2003} Cardoso V, Yoshida S, Dias O J C and Lemos J P S 2003 \textit{Phys. Rev. D} \textbf{68} 061503(R)
\bibitem{abdalla2005} Abdalla E, Konoplya R A and Molina C 2005 \textit{Phys. Rev. D.} \textbf{72} 084006
\bibitem{schen2008} Chen S and Jing J 2008 \textit{Mod. Phys. Lett. A} \textbf{23} 359


%-----------------------------------Quint-------------------------------
\bibitem{perlmutter1999} Perlmutter S et al. 1999 \textit{Astropphys. J.} \textbf{517} 565
\bibitem{riess1998} Riess A G et al. 1998 \textit{Astronom. J.} \textbf{116} 1009
\bibitem{weinberg} Weinberg S 1989 \textit{Rev. Mod. Phys.} \textbf{61} 1
\bibitem{peebles} Peebles P J E and Ratra B 1988 \textit{ApJ. Lett.} \textbf{325} L17
\bibitem{ratra} Ratra B and Peebles P J E 1988 \textit{Phys. Rev. D} \textbf{37} 3406
\bibitem{caldwell} Caldwell R R, Dave R and Steinhardt P J 1998 \textit{Phys. Rev. Lett.} \textbf{80} 1582

%%%-------------------------NHT for deSitter---------------------------
\bibitem{chambers} Chambers C M and Moss I G 1994 \textit{Phys. Rev. Lett.} \textbf{73} 617
\bibitem{bhattacharya2007} Bhattacharya S and Lahiri A 2007 \textit{Phys. Rev. Lett.} \textbf{99} 201101


\bibitem{ECSching} Ching E S C, Leung P T, Suen W M and Young K 1995 \textit{Phys. Rev. Lett.} \textbf{74} 2414; 
                   \textit{Phys. Rev. D} \textbf{52} 2118.
\bibitem{brady1992} Brady P R and Poisson E 1992 \textit{Class. Quantum Grav.} \textbf{9} 121
\bibitem{brady1997} Brady P R, Chambers C M, Krivan W and Laguna P 1997 \textit{Phys. Rev. D} \textbf{55} 7538
\bibitem{brady1999} Brady P R, Chambers C M, Laarakkers W G and Poisson E 1999 \textit{Phys. Rev. D} \textbf{60} 064003
\bibitem{molina} Molina C, Giugno D, Abdalla E and Saa A 2004 \textit{Phys. Rev. D} \textbf{69} 104013
\bibitem{NV2013} Varghese N and Kuriakose V C 2013 \textit{Gen. Relativ. Gravit.} \textbf{45} 189

%%------------------------Flat Dirac-------------------------------------------
\bibitem{jjing2004} Jing J L 2004 \textit{Phys. Rev. D} \textbf{70} 065004
\bibitem{jjing2005} Jing J L 2005 \textit{Phys. Rev. D} \textbf{72} 027501
\bibitem{jjing2006} He X and Jing J L 2006 \textit{Nucl. Phys. B} \textbf{755} 313
\bibitem{moderski2008} Moderski R and Rogatko M 2008 \textit{Phys. Rev. D} \textbf{77} 124007
\bibitem{gibbson2008a} Gibbson G W and Rogatko M 2008 \textit{Phys. Rev. D} \textbf{77} 044034
\bibitem{gibbson2008b} Gibbons G W, Rogatko M and Szyplowska A 2008 \textit{Phys. Rev. D} \textbf{77} 064024

%%------------------------dS Dirac-------------------------------------------

\bibitem{zhidenko2004} Zhidenko A 2004, \textit{Class. Quantum Grav.} \textbf{21} 273
\bibitem{zhang2009} Zhang Y et al. 2009 \textit{Chin. Phys. Lett.} \textbf{26} 030401      %Dirac          
\bibitem{wang2010} Wang C Y, Yu Z, Xing G Y and Jian-Bo L 2010 \textit{Commun. Theor. Phys.(Beijing, China)} \textbf{53} 882  %RN Derac  
\bibitem{bhattacharya2012} Bhattacharya S and Lahiri A 2012 \textit{Phys. Rev. Rev. D} \textbf{86} 084038

%-----------------------------------QNM Quint-------------------------------
\bibitem{kiselev} Kiselev V V 2003 \textit{Class. Quant. Grav.} \textbf{20} 1187
\bibitem{tharanath2013} Tharanath R and Kuriakose V C 2013, \textit{Mod. Phys. Lett. A} \textbf{28} 1350003
\bibitem{tharanath2014} Tharanath R, Nijo Varghese and Kuriakose V C 2014, \textit{Mod. Phys. Lett. A} \textbf{29} 1450057
\bibitem{birrell} Birrell N D and Davies P C 1982 \textit{Quantum Fields In Curved Space} (Cambridge, Uk:Univ. Pr.) 2nd edn.
\bibitem{HTcho2003} Cho H T 2003 \textit{Phys. Rev. D} \textbf{68} 024003
\bibitem{anderson1991} Anderson A and Price R H 1991 \textit{Phys. Rev. D} \textbf{43}, 3147
\bibitem{nollert1999} Nollert H P 1999 \textit{Class. Quantum Grav.} \textbf{16} R159
\bibitem{brady4} P. R. Brady, I. G. Moss, and R. C. Myers, Phys. Rev. Lett. {\bf 80} 3432 (1998).
\end{thebibliography}
\end{document}